\documentclass[sigconf,natbib=false]{acmart}

\usepackage{graphicx}
\pdfoutput=1
\usepackage{pgfplots}  
\usepackage{xcolor,colortbl}
\usepackage{url}
\usepackage{parskip}

\usepackage{enumitem}
\usepackage{graphicx}
\usepackage{verbatim}
\usepackage{adjustbox}

\usepackage[noend]{algorithmic}
\usepackage{algorithm}
\usepackage{amsthm}
\usepackage{lipsum}
\usepackage{parskip}

\usepackage[style=ACM-Reference-Format,backend=bibtex,sorting=none]{biblatex}
\pgfplotsset{width=5.5cm,compat=1.7}  
\setcopyright{none}

\renewcommand\footnotetextcopyrightpermission[1]{} 
\begin{document}
\title{ Trusted IP solution in multi-tenant cloud FPGA platform. }

\author{Muhammed kawser Ahmed}
\authornote{}
\email{muhammed.kawsera@ufl.edu}
\affiliation{%
  \institution{Computer Engineering, University of Florida}
  \streetaddress{}
  \city{Gainesville}
  \state{FL}
  \country{USA}
  \postcode{32603}
}

\author{Sujan Kumar Saha}
\authornote{}
\email{sujansaha@ufl.edu}
\affiliation{%
 \institution{Computer Engineering, University of Florida}
  \streetaddress{}
 \city{Gainesville}
 \state{FL}
  \country{USA}
  \postcode{32603}
}

\author{Dr. Christophe Bobda }
\email{cbobda@ufl.edu}
\affiliation{%
   \institution{Computer Engineering, University of Florida}
 \streetaddress{}
 \city{Gainesville}
 \state{FL}
 \country{USA}
 \postcode{32603}
}

\begin{abstract}
Because FPGAs outperform traditional processing cores like CPUs and GPUs in terms of performance per watt and flexibility, they are being used more and more in cloud and data center applications.
There are growing worries about the security risks posed by multi-tenant sharing as the demand for hardware acceleration increases and gradually gives way to FPGA multi-tenancy in the cloud. The confidentiality, integrity, and availability of FPGA-accelerated applications may be compromised if space-shared FPGAs are made available to many cloud tenants. We propose a root of trust-based trusted execution mechanism called \textbf{TrustToken} to prevent harmful software-level attackers from getting unauthorized access and jeopardizing security. With safe key creation and truly random sources, \textbf{TrustToken} creates a security block that serves as the foundation of trust-based IP security. By offering crucial security characteristics, such as secure, isolated execution and trusted user interaction, \textbf{TrustToken} only permits trustworthy connection between the non-trusted third-party IP and the rest of the SoC environment. The suggested approach does this by connecting the third-party IP interface to the \textbf{TrustToken} Controller and running run-time checks on the correctness of the IP authorization(Token) signals. With an emphasis on software-based assaults targeting unauthorized access and information leakage, we offer a noble hardware/software architecture for trusted execution in FPGA-accelerated clouds and data centers.
\end{abstract}

\settopmatter{printfolios=true}
	\maketitle
\pagestyle{plain}

\section{Introduction}

FPGAs are being introduced into the cloud and data center platforms for increased performance, computation, and parallelism benefits over existing accelerators such as GPUs. Technology has increased the demand for high-speed cloud computation over the last few years. Commercial and public cloud providers initiated using FPGAs in their cloud and data centers to provide tenants to customize their hardware accelerators in the cloud. The integration of FPGAs in the cloud was initiated after Microsoft published its research on Catapult in 2014\cite{Catapult}. Since then, it has become a niche technology for cloud service platforms, and major cloud provider giants, e.g., Amazon\cite{amazon}, Alibaba\cite{alibaba}, Baidu\cite{baidu},
etc., have integrated FPGAs into their platforms. For computationally intensive loads like machine learning, digital image processing, extensive data analytics, genomics, etc., users can exploit FPGA acceleration in cloud platforms 
By mentioning three concrete examples, we can express the unique advantages of FPGAs over traditional accelerators. 1. Microsoft Bing search engine experienced a 50 percent increase in throughput, or a 25 percent reduction in latency by using FPGAs in their data centers \cite{Catapult}. 2. Using Amazon AWS FPGA F1 instance, the Edico Genome project\cite{example_gnome} has over ten times the speed and performance increase for analyzing genome sequences. and 3. Introduction of Xilinx Versal FPGAs for real-time video compression and encoding in the cloud platform has significantly reduced the operating cost by reducing the encoding bitrate by 20\%, \cite{example_ai}.  

Multi-tenant FPGA platforms have multiple security concerns even after improving performance significantly. Cloud FPGAs allow tenants to drive their custom hardware designs on cloud FPGA fabric, potentially exposing multiple security risks for the adversaries, unlike CPUs and GPUs. Multi-tenancy in cloud FPGAs can create some unique hardware-related security risks such as side-channel attacks \cite{sidechannel1} where sensitive information of the hardware surface is leaked or transferred by invasive/non-invasive probing, or covert channel creation between the tenant's fabric which attacker can create a hidden channel layer between each other to transmit confidential information \cite{covert1}, 
Malicious adversaries can also launch Denial of Service (Dos) in any layer of the cloud system, including user or infrastructure level\cite{covert1},. Malicious bitstream upload into the FPGA fabric is one of the major attacks besides traditional cloud FPGA attacks, which can lead to shutdown or faulty results \cite{bitstream1}. Due to hardware fabric intervention and manipulation, short circuits fault, performance degradation, or shutdown major public cloud provides, e.g., Amazon only offers single bitstream deployment for each FPGA board. Whereas, in academics, multi-tenancy is under active research. In the case of multi-tenant cloud, FPGA security risk is more severe and intensive, as the same FPGA board fabric is shared among different tenants, which exposes the hardware surface more openly. In many reconfigurable SoCs sandboxing architecture e.g. Xilinx Zynq - 7000 SoC, UltraScale+ \cite{xilinx_trustzone}, ARM TrustZone the secure key is stored in battery-backed RAM (BBRAM) or eFuse medium. However, these methods have the following disadvantages: 
1) They still need some secure random key generation methods like random number generator (RNG) as a source of the root of trust. 2) eFuse is a fixed, non-update-able memory. 3) A physical battery is required for the BBRAM method. 4) Storing secret keys in the non-volatile memory of the cloud FPGA boards seems impracticable as the boards are rented for a specific period, and retrieving the keys from the NVM would be impossible after the tenant lease is expired. In this study, a cutting-edge method for creating identity tokens in a multi-tenant cloud FPGA platform without requiring non-volatile memory (NVM) is proposed. In order to address the aforementioned issues with the multi-tenant cloud platform, we present a brand-new, effective, and trusted solution in this paper called the \textbf{\textbf{TrustToken}}. Without employing non-volatile memory or a secure boot mechanism, we created identity tokens during runtime using an asymmetric cryptographic method. We are motivated and inspired by the Google Chromium sandboxing \cite{google_sandbox} concept to create a secure execution environment in the background of a multi-tenant platform by allocating Tokens for each IP core. To create distinct token keys, the \textbf{TrustToken} controller uses a hybrid Physical Unclonable Function (PUF) IP block. These token keys serve the untrusted IP core as an authentication ID card and must be given in each access request for a data transaction. In conclusion we state some key contributions for our protocol framework:

\begin{enumerate}[leftmargin=*] 
	\item  Provides IP isolation by   restricting IP to only allowable  interactions.
	 \item Enforces  root of trust based runtime asymmetric authorization mechanism by using a lighweight hybrid ring oscillator PUF. This approach doesnot require any non-volatile memory like eFuse or BBRAM.
	\item Efficient implementation of secure isolation by assigning secure token ID for every non-trusted IP core. 
\end{enumerate}

\section{Related Work } 
In the context of the multi-tenant FPGA platform, no prior research was found on unauthorized access and untrusted IPs security. So we will continue the discussion section with respect of reconfigurable SoC devices. In paper \cite{physicalisolation_huffmire}, Huffmire proposed the first isolation mechanism in the SoC platform. This isolation mechanism named "Moats and drawbridges" is configured by creating a fence around the untrusted IP block using a block of wires(moats). Then, a "drawbridge" tunnel is created to communicate from the fenced region. 
To prevent malicious attacks by hardware trojans, Hategekimana et al. proposed an isolation mechanism within a hardware sandbox. But the drawback of this method is only blocks specific violations. Hategekimana et al. \cite{isolation_7} presented a protocol to build secure SoC kernels of  FPGA-accelerators in heterogeneous SoC systems enforcing the Hardware Sandbox concept. The main drawback is that it has been implemented with increased overhead and latency, which seems unfit for applying real-world scenarios. Also, these proposed methods don't provide any protection outside of predefined sandboxing conditions, making the protocol ineffective in runtime conditions. 

Using Mobile SRAM PUFs, Zhao et al. \cite{Zhao_trustzone_sram} propose a prototype that extends ARM TrustZone technology to establish a root of trust-based authorization (secure storage, randomness, and secure boot). Among its disadvantages, SRAM is a poor, unreliable PUF solution and requires additional error correction code (ECC) algorithms to be applied with a helper data mechanism, increasing the post-processing cost. Using separate Hash, RSA, and AES modules, Zhao et al. proposed a two-factor authentication protocol in the paper \cite{Zhao_trustzone_token}. This work had poor implementation latency and wasn't compatible with real-world SoC IP security measures. \cite{basak_2017} describes a framework for wrapping third-party IP cores in a security wrapper and within the security wrapper,
 Security policies were added to prevent malicious activities in
the system. As a result of its high overhead and latency, this work is primarily intended for verification and debugging purposes. It is not suitable for runtime trojan mitigation or software-level attacks prevention.

Shared peripherals such as Ethernet, DRAM, UART, etc., in the ARM TrustZone architecture, are susceptible to row-hammer attacks and denial-of-service attacks\cite{pinto_arm_2019}.
Weak and inefficient authentication mechanisms are the primary security concern of ARM TrustZone technology. Several research works have reported unauthorized kernel-level privilege gain on ARM TrustZone platforms in normal world environments \cite{pinto_arm_2019}. Also, a trusted kernel region can have several vulnerabilities which can damage the whole TEE. \cite{pinto_arm_2019}. Benhani et al. \cite{benhani_trustzone}  have published a paper demonstrating several attacks on TrustZone from the simple CAD command with some simple lines of code. 

\section{Background}

\subsection{Multi-tenant Cloud FPGA}
FPGAs have mostly been used in the verification phase of ASIC designs over the previous ten years, where the ASIC design was implemented for validation and verification phases before it was actually produced. Additionally, specialized markets and research programs had some other applications. However, FPGAs are gaining popularity as an alternative for CPUs and GPUs due to high performance processing and parallelism. FPGA boards are both available on the market today as supported devices that may be connected by PCIe ports or as part of the same System-on-Chip (SoC). The integration of FPGA in cloud computing platforms to enable customers to create their hardware accelerators is one of the most recent trends in FPGA dominance. There are typically four basic methods for deploying FPGA boards in cloud data centers. As follows:
1. Coprocessor  (FPGA is coupled with CPU by PCIe cards)
2. Distinct (FPGA is deployed as a standalone component)
3. Bump-in-the-wire (FPGA is positioned between the NIC and the internet)
System-on-chip, and 4. (FPGA is fabricated in same chip die along with CPU). A cloud FPGA design known as multi-tenancy rents out the FPGA fabric to many users or tenants within the same time period. The concept of spatial silicon FPGA device sharing among several tenants is integrated into multi-tenancy. 

\subsection{Physical Unclonable Function}
In order to produce distinctive and unique ubiquitous used keys, Physical Unclonable Function uses the manufacturing variation of a silicon nanocircuit \cite{kawser_puf}. PUF can be used for a variety of cryptographic tasks, including authorization, random number generation, and authentication. The PUF theory states that even if two or more devices are identical in design, manufacturing variance will cause them to have distinct electrical properties.This variation is unpredictable and can not be estimated through observation, neither optical nor SEM. A PUF can be considered a black box, where an input challenge pair generates an output response pair. Due to manufacturing variation, the generated output response should be unique and can be used as a unique ID or authentication key. The most common existing PUF designs are Arbiter PUF, Ring Oscillator, XOR, SRAM, etc. The three most popular indicators are employed to assess the performance of PUF produced keys. They are uniqueness, randomness, and bit error rate.

\vspace{-2.5 mm }
\subsection{ARM TRUSTZONE}
In order to isolate trusted and untrusted software and hardware, ARM TrustZone technology refers to a secure execution environment (SEE)  \cite{trustzone_white}. It also goes by the name Trusted Execution Environment (TEE), and it contains a monitor that manages how these two different worlds communicate with one another. TEE TrustZone is an embedded security technology that uses two physically independent processors for the trusted and untrusted worlds. This architecture's primary flaw is the fact that similar peripherals like Ethernet, DRAM, UART, etc. are shared. Combining a few IP blocks, ARM TrustZone enables the division of groups of I/O Peripherals, Processors, and Memory into two distinct universes. Two NS bit registers on the ARM TrustZone platform are dedicated to implementing the isolation of a software process. \cite{trustzone_white}.

\subsection{Hardware Trojan and Design For Trust}

According to \cite{trojan_2}, a hardware Trojan (HT) is defined as a malicious attacker who has knowingly altered a system circuit and caused a change from intended behavior when the circuit is implemented. Because it can leak private information or alter a circuit's specifications during operation, HT poses a serious threat to SoC design. By reducing the circuit's total system functionality, HT can cause an emergency. It is exceedingly challenging to monitor this HT's negative effects during the verification stage because it is frequently deployed in stealth mode and only activated under unusual circumstances. 

\section{Threat Model and System Properties}
\label{sec:threat}
\begin{figure}[h]
	\centerline{\includegraphics[width=6cm]{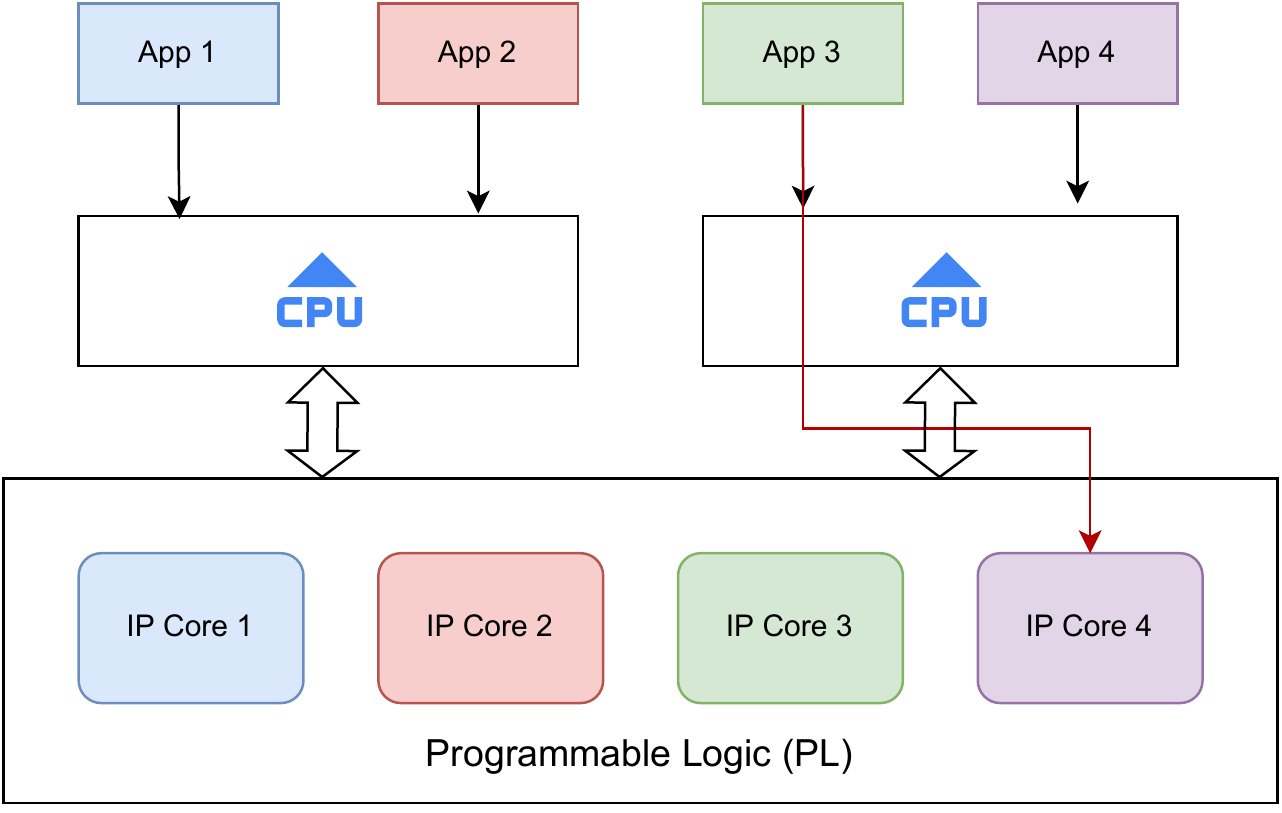}}
	\caption{Illegal software access request by Application 3 running on CPU}
	\label{fig:illegal_access}
\end{figure}

\noindent
Our threat model can be separated into two distinct explicit scenarios: Hardware Trojan and Illegal software access. Every IP is viewed as untrusted and capable of having dangerous Trojan components concealed inside it when we consider the likely first threat model. They can act under unusual circumstances. We assume that they are only used in run-time environment circumstances and are secretly carried out from the IP's internal architecture. The SoC IP integrator division is regarded to be reliable. In this case, the \textbf{TrustToken} can offer defense against unwanted data access, access control, alterations, and the leakage of any sensitive data from the IP core to the outside world.  In the second scenario, we consider a malevolent attacker who can steal, modify, or leak sensitive information by obtaining unauthorized software access from the embedded SoC world.
Figure \ref{fig:illegal_access}, as an illustration, depicts an instance of malevolent unauthorized access. In this diagram, two different CPUs that are a part of the same SoC system are powering four software-level apps.
In the hardware level design, four special IPs are added in the tenant fabric. These IPs are accessible from the software side and are identified by the same color as the related application. By using the suggested architecture model, an access request made by software Application 3 (three) to IP Core 4 (four) can be marked as unauthorized and isolated. 

The physical assaults carried out by physical apparatus were not taken into account in this essay because they fell outside of its scope. Attacks against hardware such as side-channel, probing, snooping, timing, denial-of-service, fault injection, etc. were not included in the attack scenarios. To summarize, we have considered the following threat models when describing our architecture:

\begin{enumerate}[leftmargin=*]

	\item Any malicious HT attempting to execute in runtime environments while hiding inside an IP core. We presume that concealed HT can evade detection by current CAD tools and remain undiscovered up until the payload condition is activated. 
	\item Any malevolent HT attempting to control access or transfer data without authorization. We take into account the possibility that attackers could purposefully alter the computing result by overwriting the data on a particular communication channel. We also consider that a malicious attacker could potentially cause data leakage by altering the IP core's operating mode. 
	\item Any malicious attacker trying to access other programs' sensitive data without authorization or leak it from the CPU core.
	
\end{enumerate}

\subsection{Design assumptions}
While developing our proposed solution, we have taken some key points under consideration.

	\begin{enumerate}[leftmargin=*]
\item \textbf{Multi-tenancy.} Our proposed protocol targets the multi-tenant cloud FPGA platform and observe the implementations on this platform. We assume that our proposed protocol is designed to perform in run time environment.   
\item \textbf{Adaptability.} Although this article insist on building Token based
security features for non-trusted IPs in multi-tenant cloud FPGA 
platform, this protocol can be easily adapted
only in re configurable Programmable Logic (PL) fabric-based system without the use of
processor system. 
\item \textbf{Bus Specification.} For developing our protocol, we have considered all
the interfaces established by AMBA bus specifications.
We predict that our security protocol will adopt
all necessary standard AMBA interfaces like AXI,
AHB, APB, etc. We have chosen the APB interface for
our implementation and added the necessary signals
to build our security framework.
\item \textbf{Runtime Secure Key Generation . } We considered that generated authorization tokens will be generated in runtime condition exploiting process variation and will placed in blockram inside TrustToken controller.
\end{enumerate}

\noindent
\section {Proposed Architechture }

\begin{figure*}[h!]
	\centerline{\includegraphics[width=13cm]{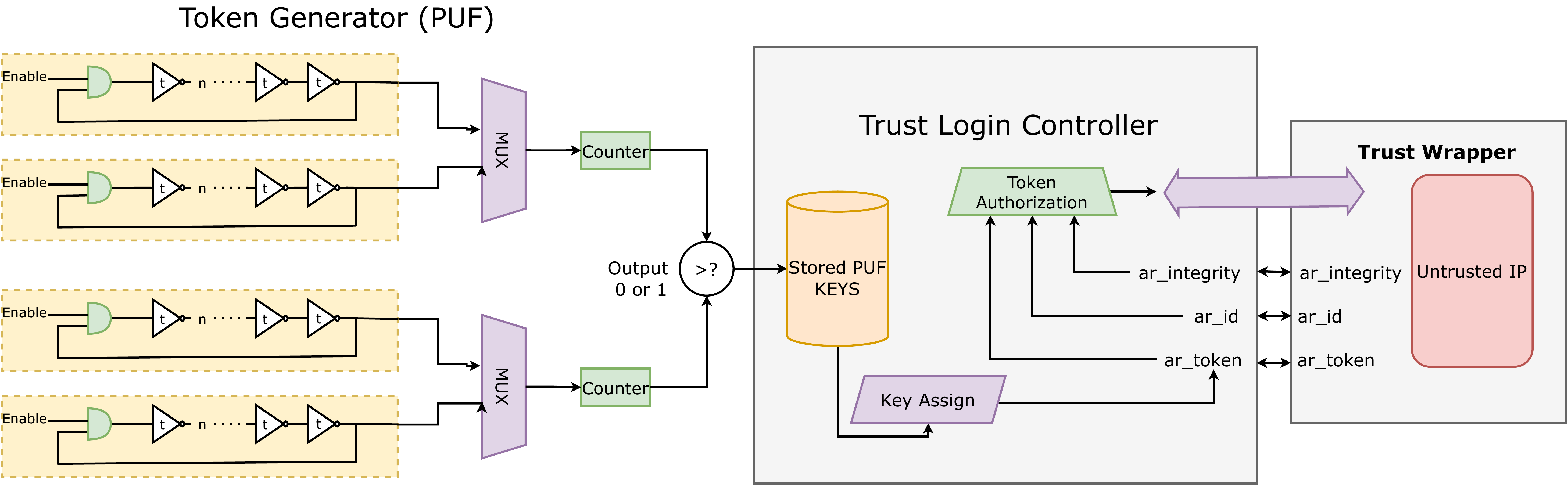}}
	\caption{Overview of the proposed TrustToken architecture framework. Consist of TrustToken Controller, TrustWrapper and TokenGenerator}
	\label{fig:architechture}
	
\end{figure*}

With the help of the TrustToken, a SoC owner will be able to offer safe and adaptable IP core communication without the need for any additional secure storage services or systems.
The TrustToken Controller, TrustWrappers, and TokenGenerator are the three parts that make up the detailed architecture of our suggested solution, which is shown in Figure \ref{fig:architechture}.

\paragraph{\textbf{TrustToken Controller.}}

\begin{figure}[h]
	\centerline{\includegraphics[width=7cm]{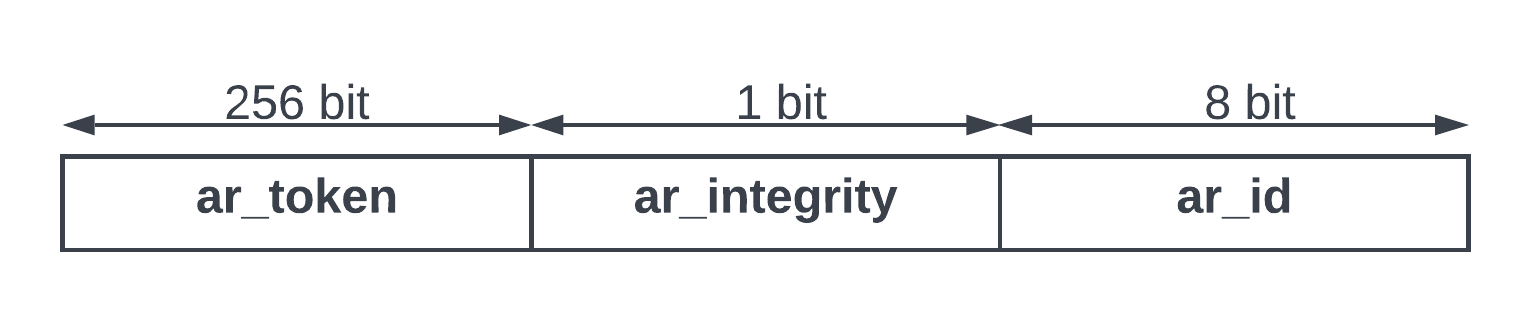}}
	\caption{TrustWrapper data ports: Proposed TrustToken signals with their relative port width.}
	\label{data_bits}
\end{figure}

 A separate centralized IP called the \textbf{TrustToken} controller is responsible for creating special Tokens/IDs for the IPs and upholding the security norms in the networked environment. To assert the validity check, any IP Integrator must modify the value of the token's parameter designated \textbf{\textit{ar\_integrity}} (Fig  \ref{data_bits}). The isolation feature will be disabled when this value is set to LOW. When HIGH, it will enforce the IP's isolation mechanism and, following a successful authorisation, execute the IP in a non-trusted zone.
The Central TrustToken Controller receives the keys once they have been generated by the PUF module and uses them to assign token IDs. The central security command center of the entire SoC system, the Central TrustToken Controller is in charge of distributing all Token IDs provided by the integrated PUF module. The \textbf{TrustToken} controller verifies the Token ID received with the list of securely stored Tokens whenever any IP wants a READ/WRITE access. It will immediately allow the data channel for communication after a successful permission or immediately disable it.

\paragraph{\textbf{Trust Wrapper.}}
\label{sub:isolation}

Every IP in our suggested design will be protected by a security wrapper called TrustWrapper. There are two distinct operating interfaces for TrustWrapper: Secured and Non-secured. The bus signals ID and Token will be added to every non-trusted IP core that is marked as non-secured. We rely on adding additional bus signals to the current AMBA bus protocol specifications in place of providing any register level isolation method or separate bus protocol for the secure isolation. It might be necessary to change the interconnect bridge mechanism for security check activities if a separate bus protocol for isolation is added. This could lead to new vulnerabilities. Additionally, in order to convey IPs ID and Token information uniformly and uniquely, a bus protocol would need to handle all conceivable bus protocol parameters, such as bandwidth, channel length, burst size, streaming techniques, etc. The Central TrustToken Controller will issue an authorization request for each data transaction started by the untrusted IP core. The controller block should receive valid and recent security data (IDs and Tokens) from non-trusted IPs via the security wrapper.

\paragraph{\textbf{Token Generator. }}
\noindent

The enhanced Ring Oscillator-based PUF, which is more stable than the original Ring Oscillator PUF, is implemented due to its low overhead and latency. Comparing Ring Oscillator-based PUF to SRAM PUF, Arbiter PUF, TRNG, or other crypto cores, the results for latency and resource consumption are encouraging.
Our unique Ring Oscillator-based PUF system can produce keys with a 256-bit width. It satisfies our requirement for offering heterogeneous SoC security by having an acknowledged uniqueness and randomization. Strong PUF is characterized as having the following security qualities in one of the fundamental studies on PUF \cite{Maes2010PhysicallyDirections}, 1. The PUF circuit cannot be physically duplicated. 2. It will enable a large number of Challenge-Response Pairs (CRPs), preventing the adversary from launching a brute force assault in a reasonable amount of time. According to the Strong PUF definition, the suggested work qualifies as a strong PUF and will be the best option to implement for the suggested SoC security justification.

\section {Proposed Protocol evaluation  }
The protocol robustness in the outlined attack scenarios was our goal in this section.

\subsection{Case 1 : ID signals being compromised}

  We described a potential attack scenario where a software level attack was introduced from an arbitrary application core in section \ref{sec:threat}. By launching a transaction request from another IP core, the malicious adversaries configure a secured IP core and try to access the victim IP. All assigned IDs and Tokens, as well as their corresponding source and destination IPs, are nevertheless recorded by Central Trust Controller. Since the attacker tried to get access illegally from a different IP core, this attempt will be checked against the saved credentials and blocked if they don't match.


\subsection{Case 2  : Insecure access control }

At the AXI interconnect level security check is carried out in the case of Xilinx TrustZone, \cite{xilinx_trustzone}, and it plays a crucial part in the security. There is a significant security risk because this Interconnect crossbar is also in charge of determining the security state of each transaction on the associated AXI bus. By altering a few security signals, any hostile attacker wishing to breach the security layer can simply control the AXI connection crossbar. This flaw was fixed with the suggested secure architecture, which imposed a strong and secure system that makes it very impossible for any access control attack to take control of the internal signals of the Central \textbf{TrustToken} Controller. With a PUF-based Token ID key, the Central \textbf{TrustToken} Controller encrypts itself, and as a result prevents any illegal use of this IP's access control.

\subsection{Case 3: Comprising INTEGRITY LEVEL}

The status of the INTEGRITY LEVEL signals determines whether any non-trusted IP is connected to the Central \textbf{TrustToken} Controller for secure isolation. Only an IP Integrator can declare the INTEGRITY STATUS at the hardware level, as was previously indicated in the thread model section. This provides defense against any CAD or RTL script attack by requiring adequate authorisation for any alteration of this signal under runtime conditions. Additionally, any malicious attacker must display their PUF-based Token ID of the untrusted IP in order to change the status of the protection level. Benhani et al. \cite{benhani_trustzone} demonstrated in their work that a malicious attacker could only significantly disrupt the Arm TrustZone AWPROT/ARPROT signal to cause a Denial of Service (DoS) disruption in the SoC. The suggested secure transition paradigm, which stipulates that a change request should additionally go into an additional authorisation layer, can be used to avoid this situation.

\section{TrustToken Implementation in multi-tenant cloud }

The experimental setup and overhead calculations needed to create our suggested architecture and assess the robustness of the suggested \textbf{TrustToken} framework are described in this section.
The primary setup involved calculating overhead and latency for data exchanges as well as effectively implementing the design.

\subsection{ Cloud FPGA Setup}
The cloud is configured on a Dell R7415l EMC server with an AMD Epyc 7251 processor clocked at 2.09GHz and 64GB of RAM. The node is running CentOS-7 with a 3.10.0 kernel. The Dell Server and the Xilinx Alveo u50 board were linked by PCIe express. Xilinx Runtime Library (XRT) was used to initiate multi-tenant domains in the FPGA and reconfigure the regions to different tenants.

\subsection{Proposed Protocol Performance}

 By implementing and synthesizing our proposed TrustToken protocol on an Alveo u50 board, we evaluated its performance. Four symmetric crypto IP cores have TrustWrappers attached around them for evaluation (AES,DES,TRNG and RSA). For the purpose of evaluating the suggested architecture model, each TrustWrapper was given a HIGH integrity state assignment. Additionally, we launched five distinct ARM-based programs to access the computational output from the crypto cores. We successfully implemented a trusted execution environment using the TrustToken concept in our implementation and tracked the outcomes. We discussed a potential software level attack scenario in section \ref{sec:threat} where a hostile attacker from Application 3 (mapped to TRNG hardware IP core) tries to create an approved access path to RSA IP core. We put this scenario into action, and the TrustToken module stopped the attack. We also created a Xilinx TrustZone enclave using the VIVADO CAD tool, based on the work Xilinx \cite{xilinx_trustzone_cad}, to compare the protocol performance with the proposed TrustToken protocol. We successfully launch a straightforward CAD Tool attack against the Xilinx TrustZone by changing the \textbf{AWPROT} signal in a runtime circumstance scenario. Similar to how the attack attempt in the suggested technique failed, the protocol's resistance to attack by CAD tools is clearly demonstrated.

\subsection{The effectiveness of the generated keys }

\begin{figure}[h]
	\centerline{\includegraphics[width=5cm]{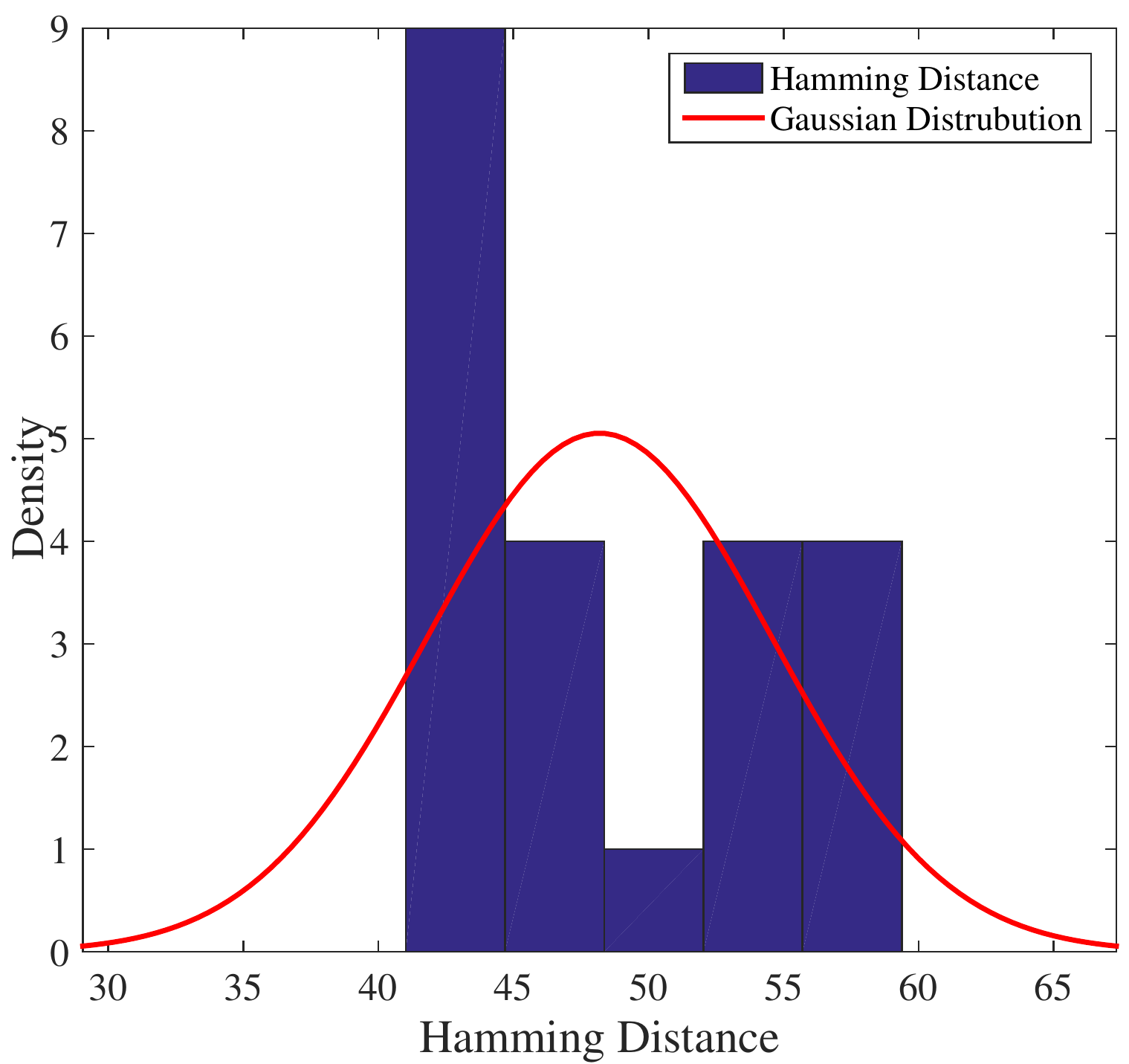}}
	\caption{Hamming Distance between the PUF keys.}
	\label{hamming}
	
\end{figure}

The results of the hamming distance calculation using the PUF keys are shown in Fig. \ref{hamming}. The hamming distance is closely rounded between 40 and 60 percent, as seen in the figure, demonstrating the keys' stability and efficacy and being extremely similar to the ideal properties of PUF \cite{kawser_puf}. The general characterizations of the PUF were compiled in Table \ref{table:PUF}. Our internal PUF architecture has 512 oscillators and is capable of producing keys that are 256 bits wide.


\begin{table}
	
	\caption{Characterizations of the Ring Oscillator PUF}
	\label{table:PUF}
	\begin{center}
		\begin{tabular}{ |c|c| } 
			\hline
			Properties  & Value \\ 
			\hline
			\hline
			Oscillators Numbers  & 512 \\
			
			Number of Keys  & 256  \\
			Single Key Length & 256 bits\\
			single Challenge Input Length  & 2 bytes\\
			
			   Randomness & 46.62\% \\
			   Uniqueness & 48.18\%\\
			   Reliability & 100\% \\
			
			\hline
		\end{tabular}
	\end{center}
	
\end{table}

\section{Conclusion }

In a multi-tenant cloud FPGA platform, this study suggests a Token-based secure SoC architecture for untrusted IP. Using a root of trust-based security module, the \textbf{TrustToken} architecture adds an extra layer of security against unwanted access control and assaults in for multi-tenant platform. The protocol can take advantage of the reconfigurable features of the SoC-based FPGA platform and uses a bespoke Ring Oscillator-based PUF module to create keys. Our method eliminates the uses of NVM memory for secure storage of keys as they are uncertain in context of multi-tenant platform.

\printbibliography

@article{catapult,
    title = {{A Cloud-Scale Acceleration Architecture}},
    author = {Caulfield, Adrian M and Chung, Eric S and Putnam, Andrew and Angepat, Hari and Fowers, Jeremy and Haselman, Michael and Heil, Stephen and Humphrey, Matt and Kaur, Puneet and Kim, Joo-Young and Lo, Daniel and Massengill, Todd and Ovtcharov, Kalin and Papamichael, Michael and Woods, Lisa and Lanka, Sitaram and Chiou, Derek and Burger, Doug},
    isbn = {9781509035083}
}

@article{example_ai,
    title = {{A Survey of FPGA-Based Neural Network Inference Accelerator}},
    year = {2017},
    journal = {ACM Trans. Reconng. Technol. Syst},
    author = {Guo, Kaiyuan and Zeng, Shulin and Yu, Jincheng and Wang, Y U and Yang, Huazhong and Wang, Yu},
    number = {11},
    volume = {9},
    arxivId = {1712.08934v3}
}

@misc{baidu,
    title = {{Baidu Deploys Xilinx FPGAs in New Public Cloud Acceleration Services}},
    url = {https://www.xilinx.com/news/press/2017/baidu-deploys-xilinx-fpgas-in-new-public-cloud-acceleration-services.html}
}

@article{pinto_arm_2019,
    title = {{Demystifying Arm TrustZone: A Comprehensive Survey}},
    year = {2019},
    journal = {ACM Comput. Surv},
    author = {Pinto, Sandro},
    volume = {51},
    url = {https://doi.org/10.1145/3291047},
    doi = {10.1145/3291047}
}

@article{xilinx_trustzone_cad,
    title = {{Design a TrustZone-Enalble SoC using the Xilinx VIVADO CAD Tool}}
}

@misc{amazon,
    title = {{Developer Preview – EC2 Instances (F1) with Programmable Hardware | AWS News Blog}},
    url = {https://aws.amazon.com/blogs/aws/developer-preview-ec2-instances-f1-with-programmable-hardware/}
}

@misc{example_gnome,
    title = {{How DNAnexus and Edico Genome are Powering Precision Medicine on Amazon Web Services (AWS) | AWS Partner Network (APN) Blog}},
    url = {https://aws.amazon.com/blogs/apn/how-dnanexus-and-edico-genome-are-powering-precision-medicine-on-amazon-web-services-aws/}
}

@misc{alibaba,
    title = {{Instance family}},
    url = {https://www.alibabacloud.com/help/en/doc-detail/25378.html}
}

@article{physicalisolation_huffmire,
    title = {{Moats and drawbridges: An isolation primitive for reconfigurable hardware based systems}},
    year = {2007},
    journal = {Proceedings - IEEE Symposium on Security and Privacy},
    author = {Huffmire, Ted and Brotherton, Brett and Wang, Gang and Sherwood, Timothy and Kastner, Ryan and Levin, Timothy and Nguyen, Thuy and Irvine, Cynthia},
    pages = {281--295},
    isbn = {0769528481},
    doi = {10.1109/SP.2007.28},
    issn = {10816011}
}

@article{kawser_puf,
    title = {{Physical Unclonable Function Based Hardware Security for Resource Constraint IoT Devices}},
    year = {2020},
    journal = {IEEE World Forum on Internet of Things, WF-IoT 2020 - Symposium Proceedings},
    author = {Ahmed, Muhammed Kawser and Yanambaka, Venkata P. and Abdelgawad, Ahmed and Yelamarthi, Kumar},
    month = {6},
    publisher = {Institute of Electrical and Electronics Engineers Inc.},
    isbn = {9781728155036},
    doi = {10.1109/WF-IOT48130.2020.9221357}
}

@misc{sidechannel1,
    title = {{Power Side-Channel Attacks on BNN Accelerators in Remote FPGAs}},
    year = {2021},
    author = {Moini, Shayan and Tian, Shanquan and Szefer, Jakub and Holcomb, Daniel and Tessier, Russell},
    arxivId = {2011.07603}
}

@article{trustzone_white,
    title = {{Programming ARM TrustZone Architecture on the Xilinx Zynq-7000 All Programmable SoC}},
    author = {{Xilinx}},
    url = {https://www.xilinx.com/support/documentation/user_guides/ug1019-zynq-trustzone.pdf}
}

@article{Zhao_trustzone_sram,
    title = {{Providing Root of Trust for ARM TrustZone using On-Chip SRAM}},
    year = {2014},
    author = {Zhao, Shijun and Zhang, Qianying and Hu, Guangyao and Qin, Yu and Feng, Dengguo},
    url = {http://dx.doi.org/10.1145/2666141.2666145.},
    doi = {10.1145/2666141.2666145}
}

@misc{google_sandbox,
    title = {{Sandbox}},
    url = {https://chromium.googlesource.com/chromium/src/+/HEAD/docs/design/sandbox.md}
}

@inproceedings{isolation_7,
    title = {{Secure Hardware Kernels Execution in CPU+FPGA Heterogeneous Cloud}},
    year = {2018},
    booktitle = {2018 International Conference on Field-Programmable Technology (FPT)},
    author = {Hategekimana, Festus and Mandebi Mbongue, Joel and Pantho, Md Jubaer Hossain and Bobda, Christophe},
    pages = {182--189},
    doi = {10.1109/FPT.2018.00035}
}

@article{basak_2017,
    title = {{Security Assurance for System-on-Chip Designs with Untrusted IPs}},
    year = {2017},
    journal = {IEEE Transactions on Information Forensics and Security},
    author = {Basak, Abhishek and Bhunia, Swarup and Tkacik, Thomas and Ray, Sandip},
    number = {7},
    month = {7},
    pages = {1515--1528},
    volume = {12},
    publisher = {Institute of Electrical and Electronics Engineers Inc.},
    doi = {10.1109/TIFS.2017.2658544},
    issn = {15566013}
}

@misc{xilinx_trustzone,
    title = {{startup • Vitis Unified Software Platform Documentation: Embedded Software Development (UG1400) • Reader • Documentation Portal}},
    url = {https://docs.xilinx.com/r/en-US/ug1400-vitis-embedded/startup}
}

@inproceedings{covert1,
    title = {{Temporal Thermal Covert Channels in Cloud FPGAs}},
    year = {2019},
    booktitle = {Proceedings of the 2019 ACM/SIGDA International Symposium on Field-Programmable Gate Arrays},
    author = {Tian, Shanquan and Szefer, Jakub},
    pages = {298--303},
    series = {FPGA '19},
    publisher = {Association for Computing Machinery},
    url = {https://doi.org/10.1145/3289602.3293920},
    address = {New York, NY, USA},
    isbn = {9781450361378},
    doi = {10.1145/3289602.3293920}
}

@article{trojan_2,
    title = {{Ten years of hardware Trojans: a survey from the attacker's perspective}},
    year = {2020},
    journal = {IET Computers {\&} Digital Techniques},
    author = {Xue, Mingfu and Gu, Chongyan and Liu, Weiqiang and Yu, Shichao and O'Neill, Máire},
    number = {6},
    month = {11},
    pages = {231--246},
    volume = {14},
    publisher = {The Institution of Engineering and Technology},
    url = {https://onlinelibrary.wiley.com/doi/full/10.1049/iet-cdt.2020.0041 https://onlinelibrary.wiley.com/doi/abs/10.1049/iet-cdt.2020.0041 https://ietresearch.onlinelibrary.wiley.com/doi/10.1049/iet-cdt.2020.0041},
    doi = {10.1049/IET-CDT.2020.0041},
    issn = {1751-861X}
}

@article{benhani_trustzone,
    title = {{The Security of ARM TrustZone in a FPGA-Based SoC}},
    year = {2019},
    journal = {IEEE Transactions on Computers},
    author = {Benhani, E M and Bossuet, L and Aubert, A},
    number = {8},
    pages = {1238--1248},
    volume = {68},
    doi = {10.1109/TC.2019.2900235}
}

@article{Zhao_trustzone_token,
    title = {{TrustTokenF: A generic security framework for mobile two-factor authentication using TrustZone}},
    year = {2015},
    journal = {Proceedings - 14th IEEE International Conference on Trust, Security and Privacy in Computing and Communications, TrustCom 2015},
    author = {Zhang, Yingjun and Zhao, Shijun and Qin, Yu and Yang, Bo and Feng, Dengguo},
    month = {12},
    pages = {41--48},
    volume = {1},
    publisher = {Institute of Electrical and Electronics Engineers Inc.},
    isbn = {9781467379519},
    doi = {10.1109/TRUSTCOM.2015.355}
}

@inproceedings{bitstream1,
    title = {{Voltage drop-based fault attacks on FPGAs using valid bitstreams}},
    year = {2017},
    booktitle = {2017 27th International Conference on Field Programmable Logic and Applications (FPL)},
    author = {Gnad, Dennis R E and Oboril, Fabian and Tahoori, Mehdi B},
    pages = {1--7},
    doi = {10.23919/FPL.2017.8056840}
}

@incollection{Maes2010PhysicallyDirections,
    title = {{Physically Unclonable Functions: A Study on the State of the Art and Future Research Directions}},
    year = {2010},
    booktitle = {Towards Hardware-Intrinsic Security: Foundations and Practice},
    author = {Maes, Roel and Verbauwhede, Ingrid},
    editor = {Sadeghi, Ahmad-Reza and Naccache, David},
    pages = {3--37},
    publisher = {Springer Berlin Heidelberg},
    url = {https://doi.org/10.1007/978-3-642-14452-3_1},
    address = {Berlin, Heidelberg},
    isbn = {978-3-642-14452-3},
    doi = {10.1007/978-3-642-14452-3{\_}1}
}

\end{document}